\documentclass[12pt]{article}
\usepackage[english]{babel}
\usepackage[utf8x]{inputenc}
\usepackage{amsmath}
\usepackage{graphicx}
\usepackage{etoolbox}
\usepackage{changepage}
\usepackage{titlesec}
\usepackage[parfill]{parskip}
\usepackage[margin=1in]{geometry}
\usepackage{times}
\usepackage{float}
\usepackage[numbers,super]{natbib}
\usepackage{lipsum} % Package to generate dummy text throughout this template. Remove for real use.

\titleformat*{\section}{\normalsize\bfseries} % Makes section titles 12 pt font

\usepackage[table]{xcolor}
\RequirePackage{fix-cm}
\usepackage{graphicx}
\usepackage{hyperref}
\usepackage{xspace}
\usepackage{color}
\usepackage{listings}
\usepackage{acronym}
\usepackage[T1]{fontenc}
\usepackage{lmodern}
\usepackage{multirow}
\usepackage{todonotes}
\usepackage{url}
\usepackage{tcolorbox}
\usepackage{comment}
\usepackage{colortbl}
\usepackage{makecell}

\usepackage{amsmath,amssymb,amsfonts}
\usepackage{algorithmic}
\usepackage{textcomp}
\usepackage{wrapfig, blindtext}
\usepackage{moreverb,url}

\newcommand\BibTeX{{\rmfamily B\kern-.05em \textsc{i\kern-.025em b}\kern-.08em
T\kern-.1667em\lower.7ex\hbox{E}\kern-.125emX}}

\acrodef{ADD}{Attribute-Driven Design}
\acrodef{QA}{Quality Attribute}
\acrodef{UML}{Unified Modeling Language}
\acrodef{IoT}{Internet of Things}

\PassOptionsToPackage{table}{xcolor}

\definecolor{lightgray}{rgb}{.9,.9,.9}
\definecolor{white}{rgb}{1,1,1}
\definecolor{darkgray}{rgb}{.4,.4,.4}
\definecolor{darkgreen}{rgb}{0, 0.39, 0.00}
\definecolor{Gray}{gray}{0.7}

\title{\large{Self-Confidence of Undergraduate Students in Designing Software Architecture}} % using \large makes the title approximately 14 pt.
\author{\normalsize Lotfi ben Othmane and Ameerah-Muhsina Jamil\\
\normalsize \{othmanel, amjamil\}@iastate.edu\\
\normalsize Iowa State University}

\date{} % This leaves the date blank.

\makeatletter % This gets the margins for the title set.
\patchcmd{\@maketitle}{\begin{center}}{\begin{adjustwidth}{0.5in}{0.5in}\begin{center}}{}{}
\patchcmd{\@maketitle}{\end{center}}{\end{center}\end{adjustwidth}}{}{}
\makeatother

\begin{document}
\raggedright
\maketitle
%\thispagestyle{empty}
%\pagestyle{empty}
%

%\author{\IEEEauthorblockN{Lotfi ben Othmane
%\IEEEauthorrefmark{1}, Ameerah-Muhsina Jamil\IEEEauthorrefmark{1}}
%\\
%\IEEEauthorblockA{\IEEEauthorrefmark{1} Iowa State University, Ames, IA USA}%}

%}

%\begin{abstract}
\section*{Abstract}
 Software architecture students, often, lack self-confidence in their ability to use their knowledge to design software architectures. This paper investigates the relations between undergraduate software architecture students' self-confidence and their course expectations, cognitive levels, preferred learning methods, and critical thinking. We developed a questionnaire with open-ended questions to assess the self-confidence levels and related factors, which was taken by one-hundred ten students in two semesters. The students answers were coded and analyzed afterward. We found that self-confidence is weakly associated with the students' critical thinking and independent from their cognitive levels, preferred learning methods, and expectations from the course. The results suggest that to improve the self-confidence of the students, the instructors should work on improving the students' critical thinking capabilities. 
 
\section{Introduction}

Undergraduate students are expected to step directly into software developer positions and succeed. Typical undergraduate students are, however, not prepared for the ambiguity of the industry~\cite{Mow1979}. The lack of self-confidence makes them resistant to take opportunities and lead projects, and their capabilities are sometimes below the expectations of the employers~\cite{HeMy2018}.  \emph{Self-confidence}, aka self-efficacy, perceived ability, and perceived competence, is a measure of one's belief in their ability to successfully execute a specific activity~\cite{Sho2010, bandura1986social, NAP2303}. According to Bandura, the outcomes that people expect depend heavily on their self-confidence that they can perform the skill~\cite{NAP2303}. 

Self-confidence was considered a critical factor that impacts undergraduate students' abilities in programming~\cite{Hanks2006,RLW2004}. For instance, Heggen and Meyers~\cite{HeMy2018} studied students' confidence before joining a program to develop real-word applications. They found that only 25\% of the students were optimistic about their abilities in developing software systems before joining a pair-programming program and are far more confident in their leadership abilities after finishing the program. Hanks also measured their students' confidence after practicing with pair-programming and found that the confident students liked pair-programming the most, while the least confident students liked it the least~\cite{Hanks2006}.

 %Researchers studied the decision-making process of software architects~\cite{HeAv2011, TRPH2017, VLV2007} and developed education programs that train the students on using the design thinking process, identifying the project's constraints and assumptions, developing candidate solutions, and evaluating them considering their advantages, disadvantages, and trade-offs. The results have been promising~\cite{VSC2017, HeAv2010, SWTF2006}.

Software architects gain cumulative architectural knowledge through experience; they make architectural decisions in ambiguous situations and learn by assessing the impacts of these decisions on software~\cite{HeAv2011, TRPH2017, VLV2007}.
Teaching software architecture is challenging given the nature of software architecture and the characteristics of the learners \cite{GaAn2016}. For instance, software architecture is a fuzzy concept, challenging to present as a tangible and useful concept to non-experienced software engineers while the learners are used to topics where the problems and solutions could be precisely defined, which do not apply to the case of architecture.

%, (2) is a practical topic where teaching activities require a proper context and problem of sufficient complexity, and (3) deals with no clear problem description, no exact solutions, and mostly team rather than individual work. Besides, the learners (1) , (2) have limited knowledge and experience to perceive the need to choose among a set of alternatives and address conflicts, (3) used to concrete concepts such as code, and (4) have limited ability to deal with abstract concepts.

Software architecture students, like programming students, have, often, a self-confidence problem. For example, some of our students expressed, in Spring 2017, that they are not able to use their knowledge to design software architectures. The problem of self-confidence of software architecture students has been addressed, in our opinion, by focusing on practicing with design patterns~\cite{RuCh2015} or adopting the clinical mode~\cite{MIRW2000}. 

We conducted informal meetings with colleagues to assess the factors that may impact the self-confidence levels of software architecture students. The goal was to identify the basic aspects that we could act on to improve the students' confidence levels. The consultation led to the selection of variables: course expectations, cognitive levels, preferred learning methods (e.g., passive, active), and critical thinking. 

%We redesigned in Fall 2017 our software architecture course to focus on the \emph{process} of designing software architecture rather than on the knowledge of architectural concepts. We adopted the \ac{ADD} process as described by Cervantes and Kazman~\cite{CeKa2016}. The students used the case studies available in the book. They practiced with the process activities in group assignments on an \ac{IoT} project and in individual assignments on Web application projects. 

We developed a questionnaire with open-ended questions to study the relationships between students' self-confidence and their expectations, cognitive levels, preferred learning methods, and critical thinking. We gave the questionnaire to the students who took the course in two subsequent semesters: Fall 2017 and Spring 2018. In total, 110 students out of 138 students took the survey. We coded the answers of each student using the descriptive coding method~\cite{saldana2015coding}, and used the frequency technique as in-text analytics~\cite{Meh2006,GKT2019} to assess the dependency between the students' self-confidence levels and their expectations, cognitive levels, preferred learning methods, and critical thinking.

The paper is organized as follows. Section~\ref{sec:relwork} discusses related work. Section~\ref{sec:crsdesign} describes the course design. Section~\ref{sec:resmethod} describes the research method. Section~\ref{sec:resultexplore} explores the collected data. Section~\ref{sec:resultsanalysis} analyses the relationships between self-confidence and expectations, cognitive levels, preferred learning methods, and critical thinking. Section~\ref{sec:limitations} discusses the impacts and limitations of the study and Section~\ref{sec:conclusions} concludes the paper.

\section{Related Work}\label{sec:relwork}

This section reports about existing work on exploring ways to teach software architecture.

Valentim et al.~\cite{VSC2017} performed a study with 17 postgraduate students on student perceptions of applying design thinking to design mobile applications. The students appreciated the process as they find it useful. However, they find it challenging to apply because they need to think creatively and generate ideas. Besides, they found the application of the techniques (e.g., workshops and brainstorming) useful but challenging given the lack of team connection and critical thinking~\cite{Hey2018}.

Heesch and Avgeriou~\cite{HeAv2010} surveyed 22 undergraduate software engineering students in the Netherlands, aiming to find out the natural reasoning process during architecting. They found that most of the students tried to understand and consider the architectural drivers and emphasize the quality attribute requirements. However, many students did not identify the most challenging requirements nor prioritize them. In addition, most of the students affirmed that they used the requirements to identify design options and preferred well-known solutions rather than unknown alternatives. They also found that while more than half of the students affirmed that they considered the pros and cons of alternative solutions, many did not consciously make trade-offs between requirements.

Schriek et al.~\cite{SWTF2006} propose a card game to help novice designers design reasoning.\footnote{Design reasoning means using logic and rational thinking to make decisions.} The cards represent the reasoning techniques: problem structuring, option generation, constraint analysis, risk analysis, trade-off analysis, and assumption analysis. The authors evaluated their technique's efficacy using twelve groups of students who took the software architecture course. The study showed that the cards trigger reasoning and lead to more discussion and reconsideration of previous decisions. The groups who used the card game identify more distinct design elements and spend more time reasoning with the design. 

Rupakheti and Chenoweth experimented with teaching undergraduate students software architecture for a decade~\cite{RuCh2015}. They found that teaching the topic is challenging because it contrasts the students' habits in the other computer science courses. For instance, software architecture requires addressing problems in large and complex software, use multiple complex solutions, and is designed from incomplete information. The authors described how they evolved the course from lecture-heavy to a hands-on course that teaches the students how to use architecture patterns to address \acp{QA} in lab experiments. The authors found that the use of labs reinforced the students learning. 

Ali and Solis~\cite{ALI201423} studied the perception of master students on the easiness of use, usefulness, and willingness to use the \ac{ADD} method in the future. They found that the students find the architecture design method useful but not easy to use and are neutral in term of willingness to use the \ac{ADD}.

Ben Othmane and Lamm~\cite{OtLa2019} studied the factors associated with the mindsets of software architecture students. They found that students' mindset weakly correlates with their cognitive levels and is related to their expectations. They also found that the students who prefer practicing software architecture have more open mindsets than those who prefer quizzes.

We did not find studies on the self-confidence of undergraduate students to design software architecture--recall that the issue has been investigated for programming students~\cite{Hanks2006,RLW2004}. We initiate the discussion about measuring the students' self-confidence and assessing the factors that may impact it. Recall that this trait is essential for students to take the initiative and lead projects. 

\section{Course Description}\label{sec:crsdesign}

The course Software Architecture Design is an undergraduate-level course for software engineering and computer engineering programs. Before taking the class, the students take a class on developing web applications. The course is given two times a year. Each semester, the class meets two times a week for 14 weeks, each of 75 min.

\begin{comment}

\begin{table}[tb]
\caption{Modules for the software Architecture design course.}
\label{tab:modules}
\centering
\begin{tabular}{p{.05in} p{1.8in}  p{0.3 in} p{.45in} } 
\hline
\rowcolor{Gray}\hline
ID& Module & \# of sessions & Teach. form\\ \hline
\rowcolor{lightgray}
1 & Overview of software architecture& 2 & R, GA \\\hline
2 &\ac{UML}& 3 & R, GA, IA \\\hline
\rowcolor{lightgray}
3 & Architecture styles, patterns, and tactics& 4 & L, R, GA\\\hline

4 & Architecture drivers& 5 & R, PA, GA, IA\\\hline
\rowcolor{lightgray}
5 & Architecture design process&4& R, GA, PA, IA\\\hline

6 & Documenting a software architecture&1& R, GA \\\hline
\rowcolor{lightgray}
7 & Architecture evaluation& 2& R, GA, IA \\\hline

8 & Software security architecture&1& L\\\hline
\rowcolor{lightgray}
9 & Architecture recovery&1&L \\\hline\hline

\end{tabular}

\begin{flushleft}The notation is: Reading (R), Group Activities (GA), Lecture (L), Individual Assignments (IA), and Project Assignment (PA).\end{flushleft}

%\vspace{-0.3in}
\end{table}

\end{comment}

%\subsection{Course objectives}

The goal of the course is to train the students in designing software architecture. The course uses the \acf{ADD} method~\cite{CeKa2016}. The students acquire the knowledge needed to design software architecture and learn how to apply the \ac{ADD} method, which is a process-based approach to the design of software architecture~\cite{HKNO2005, CeKa2016}. The objectives are:

\begin{enumerate}
\item understand and explain the importance of software architecture,
\item understand the relationships between software quality attributes and software architecture,
\item Gain ability to elicit software architecture drivers,
\item Understand the roles of a set of architecture styles, patterns, and tactics in software architecture,
\item Apply the attribute-driven method to design and evaluate software architecture.
\end{enumerate}

\begin{comment}

The course objectives were refined to identify the course modules listed in Table~\ref{tab:modules}. The table provides for each module the number of sessions and the teaching form.

\begin{table}[tb]
\caption{List of assignments.}
\label{tab:assignments}
\centering
\begin{tabular}{p{0.9in}p{2.1in} } 
\hline
\rowcolor{Gray}\hline
Type	& Description\\ \hline
\rowcolor{lightgray}
Individual assignment 1 &   Practice with UML diagrams.	\\\hline
Group assignment~1 &   Experiment with the identification of architectural drivers for a given project.\\\hline
\rowcolor{lightgray}
Group assignment~2&   Experiment with the design of an architecture for a given project.\\\hline
Individual assignment~2&  	Experiment with the design of an architecture for a given project.	\\\hline
\rowcolor{lightgray}
Individual assignment~3 &	Evaluate the architecture of a software	\\\hline
Group assignment~3 &	Implement a designed architecture and reflect about the gap between as-designed and as-implemented architecture.\\ \hline\hline
\end{tabular}
\vspace{-0.2in}
\end{table}

We applied in this course the Team-based learning method~\cite{SORM2014}. The students were given book chapters and papers, selected from~\cite{CeKa2016,Gor2006,Fow2000,Ebe2017}, to {\tt read} before coming to class. The students take an online quiz to test their readiness to discuss the topic. The quiz questions are discussed afterward in-class. As exception, modules eight and nine were designed as lectures because they are considered as complementary.

\end{comment}

The students work in groups on in-class activities. The activities include answering questions that need reflection, working on exercises, and simulating architecture meetings. The case studies provided by~\cite{CeKa2016} were useful for the students to see the use of the techniques.

The students were requested to practice the knowledge that they acquire in the lecture sessions on group and individual assignments. The students work in groups on projects in three group assignments: gathering architectural drivers, designing the architecture of the new version of a given software and implementing the architecture they designed. The individual assignments enforce the experience that the students obtained from the project. The group assignments are related to an \ac{IoT} project, while the individual assignments are related to IT projects. This is expected to give the students an experience with the two domains. 

%In addition, few sessions were dedicated to extra activities: visits of professionals from the industry, experience-sharing panel session that includes a representative of each team, and a research talk by the instructor.

\section{Research Method}\label{sec:resmethod}

The best solution to assess the relationships between student's self-confidence level and expected dependent variables (course expectations, cognitive levels, preferred learning methods, and  critical thinking)  is to specify a set of closed questions (e.g., using Likert scale and variable categories) and use inference statistics techniques. Since, we do not know the different categories for each of the dependent variables, we conducted a qualitative study. The study uses students' free-text responses to a questionnaire as the data source. We discuss the preparation of the study, the data collection, and the data analysis activities. 

%\todo[inline]{grades of the exams and assignments depend on the easiness of the questions/problems. We did not use them in the study.}

\vspace{.1in}
\noindent {\bf Preparation of the study.} We discussed the course with colleagues and identified a set of factors that we expected to be associated with students' self-confidence, which are: (1) course expectations, (2) cognitive levels by the students, (3) preferred learning methods and (4) critical thinking. Therefore, we used expert opinions rather than literature review to identify the factors that may impact the students' self-confidence in designing software architecture. The factors were used to develop a set of questions to measure them, listed in Table~\ref{tab:questionaire}. 

\begin{table}[tb]
\caption{Questionnaire.}
\label{tab:questionaire}
\centering
\begin{tabular}{p{0.1in}p{1.1in}p{4.7in} } 
\hline
\rowcolor{Gray}\hline
ID & Factor & Question\\ \hline
\rowcolor{lightgray}
1& Expectation & What was your expectation of the course before taking it?\\\hline
2& Cognitive level & Assume you are given a project and asked to design an architecture for it. How would you do the design?\\\hline
\rowcolor{lightgray}
3& Self-Confidence& How much confidence would you have about your design?\\\hline
4& Critical thinking  & What are the differences between designing the architecture of a Web application and the one of an \ac{IoT} system?\\\hline
\rowcolor{lightgray}
5& Preferred learning method & What is/are the method(s) that helped you better learn software architecture? \\\hline\hline
\end{tabular}
\vspace{-0.2in}
\end{table}

We developed an anonymous, electronic questionnaire using Google Form and made it available online for the students in November 2017 (for Fall 2017 semester) and April 2018 (for Spring 2018 semester).\footnote{The project was granted an IRB exemption.} (The students answer the questionnaire at the  end of the semester.) The submissions were anonymous, but the students had to tell the instructor that they participated in the study to get their bonus points.

\vspace{.1in}
\noindent {\bf Data collection.} One-hundred ten students answered the questionnaire in Fall 2017 and Spring 2018. We used the thematic analysis~\cite{saldana2015coding} method to extract insights from questionnaire responses. The thematic analysis approach is a method for identifying, analyzing, and reporting patterns (themes) within data~\cite{BrCl2006}. It allows exploring phenomena through interviews, stories, or observations~\cite{Con2010}. 
First, we read all the answers to the questions and extracted the thematic code representing each of the answers. A code is a word or short phrase identifying the essence of a portion of language-based or visual data~\cite{BrGr2007}. At the end of this step, we assigned codes to each of the one-hundred-ten students' responses and obtained a set of categories for each of the factors of Table~\ref{tab:questionaire}. We removed the records of eight students (and used the records of 102 students) because their answers to the self-confidence question were not clear/conclusive.

The cognitive levels of the students according to Bloom taxonomy~\cite{GaKAC2000} are commonly assessed either using test questions or reflection write-ups~\cite{HBF2019}. We used in this study the verbalization\footnote{See for example: https://adp.uni.edu/documents/ bloomverbscognitiveaffectivepsychomotor.pdf} used by the students in their responses to (reflection) Question 2 to identify the cognitive level of each student. The association of the verbs to the different levels is based on the author's domain knowledge. For instance, Participant {\tt(P20)} said \emph{"The design would vary depending on what the project requirements and architectural drivers were. Once I decided on an optimal reference architecture, I would go through the iteration design process and make sure that appropriate design decisions were made to address every architectural driver that was identified in the project description."} The codes extracted from the statements are: apply the design process, select reference architecture, and evaluate. Since the code "evaluate" is classified in the cognitive levels as {\tt Evaluation}, we ranked the student at level {\tt Evaluation}--that is, the code associated with the higher cognitive level is selected.

Next, we counted the frequencies of the different codes/categories/levels used in the responses to each of the questions of Table~\ref{tab:questionaire} and observed the patterns in these data. We discuss the data that we collected in Section~\ref{sec:resultexplore}.

\vspace{.1in}
\noindent {\bf Data analysis.} We represented the relationships between the students' self-confidence levels and each factor affecting their self-confidence using matrices--we use one matrix for each factor. The columns of a matrix are the self-confidence levels and the rows are the codes/code-categories of the factor being studied. The elements are the frequencies of the students who belong to the given factor category and given self-confidence level. We use Rao-Scott adjusted~\cite{rasc1987} Chi-square independence test~\cite{Cochran1952} to evaluate the dependencies between self-confidence and the related factors. In addition, we used the Cramer V to check the association levels of the factors.

\begin{table}[bt]
\centering
\caption{Codes used to express self-confidence of the students in their architecture designs.}
\label{tab:Confidence}
\rowcolors{2}{white}{lightgray}
\begin{tabular}{p{0.1in}p{1.2in}p{4.6in} }
\hline
\rowcolor{Gray}\hline
 ID& Confidence level	& Codes \\\hline

1 & Confident & very confident, confident, pretty confident \\\hline

2 & Moderate & somewhat confident, moderate, decent, somewhat confident, relative, quite confident \\\hline

3 & Fair & fair confidence, not very/extreme-ly/overly confident \\\hline

 4 & No confidence &not confident, not great  \\\hline

\hline
\end{tabular}
\vspace{-0.2in}
\end{table}

\section{Data Collection}\label{sec:resultexplore}

This section summarizes the responses of the students to the questionnaire and discusses the results.

\begin{figure}[tbp]
    \centering
    \begin{minipage}[t]{0.48\textwidth}
    \centering
    \includegraphics[width=\textwidth]{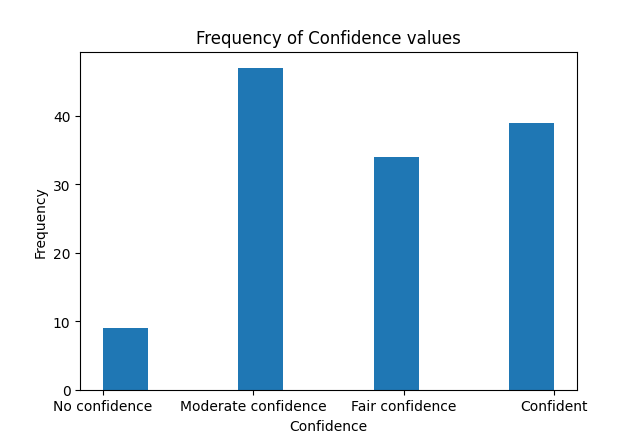}
    \caption{Frequency of self-confidence levels.}
    \label{fig:Confidencelevel}
        \end{minipage}
        ~
    \begin{minipage}[t]{0.48\textwidth}
  
    \includegraphics[width=\textwidth]{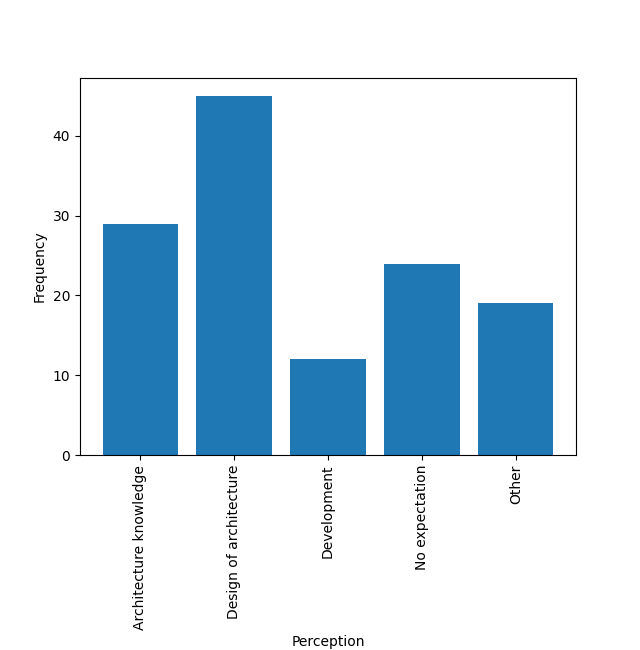}
    \caption{Frequency of expectations.}
    \label{fig:exepctations}
  \end{minipage}
\end{figure}

\subsection{Self-confidence}

This subsection discusses the results of the analysis of the responses to the question: {\tt  How much confidence would you have about your design?} We classified the extracted codes into five categories: high, moderate, fair, and no self-confidence, in addition to no definite answer category.  Table~\ref{tab:Confidence} shows the codes that we used for each level, and Figure~\ref{fig:Confidencelevel} shows the frequency of these levels. The number of students who have high self-confidence level is 34 (33\%). These students seems to be comfortable applying design processes, such as \ac{ADD}. For instance, student {\tt(P24)} expressed that by saying that \emph{"I feel I would be very confident in my design because I think the design process does a good job of ensuring the architecture considers and satisfies all the drivers. So as long as I am successful in compiling a thorough list of drivers, I think the design will turn out well."}. Out of the remaining students, we see that eight students (about 7\%) did not provide definitive answers. Students who elaborated their answers expressed the need for references to ensure the efficacy of their designs. The records of the students who did not provide a definite answer to the question are removed from the dataset.

%For instance, student {\tt(P32)} said \emph{"I don't really know, for everything we covered there were a lot of different views, but we didn't really know which one the best was. And a thing that bothers me a lot is how can we know that the QAs [Quality attributes] are addressed? Ok, I created an architecture, but I still don't know if the response time is less than 1 second or not. We have not done any testing, any coding..."}

\subsection{Student' expectations about the course}

%For instance, {\tt(P42)} said \emph{"Initially I assumed the course would cover what effectively boils down to self-congratulatory discussion filled with buzz-words. The software architectures I have seen through working for software development companies are often a chaotic mess of anti-patterns and individuals attempting to defend decisions by calling them architectural drivers. Before the course even began this was what I expected to experience."}

 This subsection discusses the analysis results of the answers to the question: {\tt What was your expectation of the course before taking it?} Figure~\ref{fig:exepctations} shows the frequency of the students' expectations about the course. In general, most of the students expected the course to be about the design of architecture (44.11\%), architecture knowledge (28.4\%), and development (11.7\%). We observe that some students related the course to other courses or to experiences they had in their internships. We also observe that the number of students who did not have a clear expectation about the course is 24 (23.5\%). The reason for this high percentage is possibly due to the fact that the course is required for their programs. Note that some students specified more than one course expectation category. (We have 129 records for 102 students.)

\subsection{Cognitive levels}

\begin{figure}[tbp]
    \centering
  \begin{minipage}[t]{0.48\textwidth}
    \includegraphics[width=\textwidth]{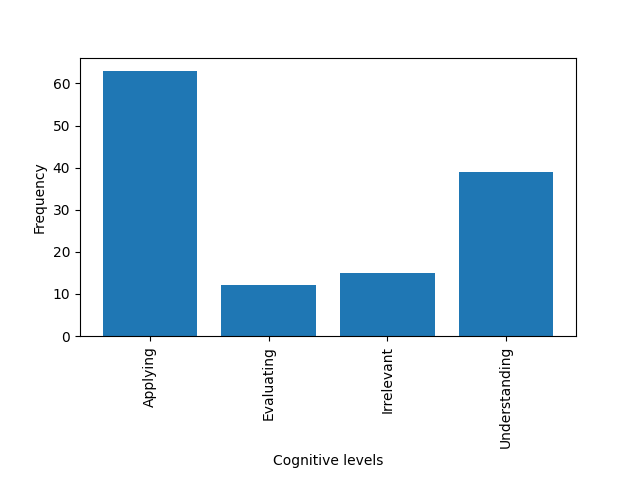}
    \caption{Frequency of the cognitive levels.}
    \label{fig:CogntiveLevel}
    \end{minipage}
    ~
    \begin{minipage}[t]{0.48\textwidth}
    \includegraphics[width=\textwidth]{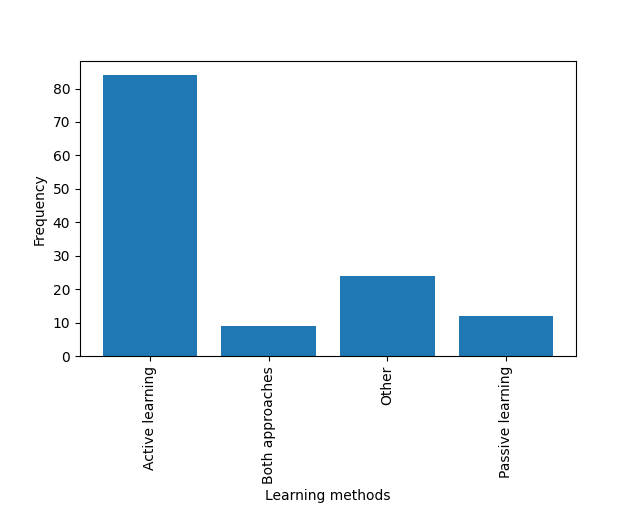}
    \caption{Frequency of learning methods.}
    \label{fig:learningmethods}
  \end{minipage}
\end{figure}

This subsection reports the results of the analysis of the replies to the question: {\tt Assume you are given a project and asked to design an architecture for it. How would you do the design?} We coded the responses of the students and classified the extracted codes based on the new Bloom cognition levels~\cite{GaKAC2000}. To comply with the conditions to use the Chi square independence tests, we merged the categories Creating, Evaluating, and Analyzing into category Evaluating and merged the category Remembering with the category Understanding. Table~\ref{tab:BloomLeaning} shows the classification of the codes to modified Boom's cognition categories, and Figure~\ref{fig:CogntiveLevel} provides the frequency of the cognitive levels. 

\begin{table}[bt]
\caption{Cognitive levels of the students.}
\label{tab:BloomLeaning}
\centering
\rowcolors{2}{white}{lightgray}
\begin{tabular}{p{1.1in}p{4.9in} } 
\hline
\rowcolor{Gray}\hline
 Level & Codes\\ \hline
 Creating & (Combined with Evaluating)\\ \hline
 Evaluating & identify trade-offs, identify risk, architecture evaluation, adjust design process\\ \hline
 Analyzing & (Combined with Evaluating) \\ \hline
 Applying & identify architecture drivers, get requirements, meet stakeholders, create design, apply the design process, do as in assignments, modify reference architecture \\ \hline
 Understanding & select reference architecture, select architecture style, select architecture type, make diagrams \\ \hline
  Remembering	& (Combined with Understanding)\\ \hline
 Irrelevant	&	 \\ \hline \hline

\end{tabular}
 \vspace{-.1in}
\end{table}

We observe that most of the students (46.07\%) have "applying" cognitive level and that 13.7\% of students provided irrelevant answers. Many of these students specified that they need more details to decide how to proceed with the design or provided non-useful answers such as \emph{"I would probably try and layout the entire system's architecture in one go because the process of iterations confused me."} {\tt(P11)}. We note that only few students (8.8\%) had the evaluating cognitive level.   

\subsection{Preferred learning methods}

This subsection discusses the results of the analysis of the answers to the question: {\tt What is/are the method(s) that helped you better learn software architecture?}. We grouped the preferred methods into passive methods, active learning methods, passive and active methods, and other methods. The other category includes, for examples, students who did not specify a definitive learning methods or students who referred to other courses, internet, etc. Figure~\ref{fig:learningmethods} shows the frequency of preferred learning methods--a student can specify multiple methods. We observe that most of the students prefer active learning methods, that is 65.1\%.

\subsection{Critical thinking}

\begin{table}[btp]
\caption{Codes associated with the critical thinking aspects.}
\label{tab:cabsynthesis}
\centering
\rowcolors{2}{white}{lightgray}
\begin{tabular}{p{1.2in}p{4.9in} } 
\hline
\rowcolor{Gray}

Aspect & Codes \\\hline
		
 Architecture drivers &  reliability, interoperability, scalability, architecture drivers, availability, performance, and security requirements \\ \hline

 Patterns of the structures of the solutions & communication pattern, components structure (e.g., modularity), control of physic, Objects vs logic computation, interacting actors, access to arch. components, flexibility to add components, integration of complex software, simplicity and complexity, technology stack, security protocols, use of hardware, complexity of software, configuration management  \\ \hline
 
 Architectural knowledge & Reference architecture, architecture styles, architecture patterns\\ \hline

 No definite answer & \\\hline\hline
\end{tabular}
\vspace{-0.2in}
\end{table}

We assess students critical thinking by evaluating their abilities to identify the differences between the architectures of Web applications and IOT-based software. This subsection reports the results of the analysis of the replies to the question: {\tt What are the differences between designing the architecture of a Web application and the one of an IoT system?} Table~\ref{tab:cabsynthesis} provides the codes that we derived from the responses, which are classified into four categories: architecture drivers, patterns of the solutions' structures, architectural knowledge, and no definite answer. Note that some students identified difference in more than one category; i.e., a student could discuss performance, which is an architecture driver, and distribution of the system's components, which is a pattern of the solution' structure.

\begin{wrapfigure}{r}{0.5\textwidth}
\vspace{-.5in}
    \includegraphics[width=.5\textwidth]{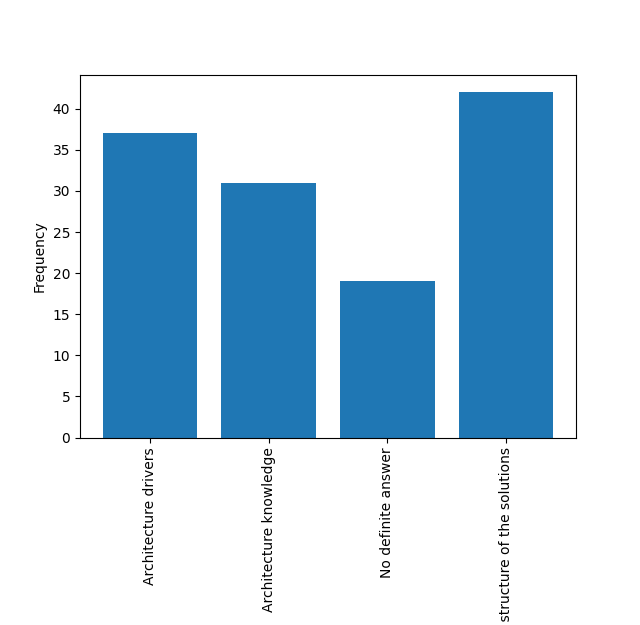}
    \caption{Frequency of critical thinking aspects.}
    \label{fig:criticalthinking}
    \vspace{-.4in}
\end{wrapfigure}

Figure~\ref{fig:criticalthinking} provides the frequency of the critical thinking aspects. We observe that the number of students who expressed that the two types of systems use different architecture structure patterns is the largest, 42 (32.5\%) and the number of students who did not provide definitive answers is 19 (14.7\%). Some of these 19 students reported "they do not know", did not answer the question, or provided non-useful answers such as \emph{"I thought this was a survey, not a test."}. This is a good results as the students are expected to have limited experience with the technology stack but are expected to reason about the architecture drivers, patterns, and tactics. 

\section{Analysis of the relationships between self-confidence and course expectations, cognitive levels, and preferred learning methods, and critical thinking}\label{sec:resultsanalysis}

In this section, we analyze the relationships between students' self-confidence levels and their expectations, critical thinking, cognitive levels, and preferred learning methods. We use in this analysis the Chi-square independence test~\cite{Cochran1952} with Rao-Scott adjustment~\cite{rasc1987} and the items frequencies. 

\subsection{Students' course expectations}

Table~\ref{tab:PerceptConfidence} provides the frequencies of the course expectations vs. self-confidence levels of the students. We observe that most of the students who have no expectations are either confident (8 students) or have moderate confidence (8 students). While course instructors may not have control over students' expectations when coming to the course, they should ensure that the students have correct expectations (i.e., the course is about architecture design) when the course starts. Having one question about course expectations in a course welcoming survey quiz--that could be administered in the first seasons of class--would help fix the students' expectations.

\begin{table}[btp]
   \caption{Students' course expectations vs self-confidence levels.}
\label{tab:PerceptConfidence}
\centering
\rowcolors{2}{white}{lightgray}
\begin{tabular}{p{2.0in}p{0.9in}p{0.9in}p{0.9in}p{0.9in}p{1.0in} } 
\hline
\rowcolor{Gray}
Perception              &Confident &Fair confidence &Moderate confidence &No confidence\\\hline
  Architecture knowledge &      7               &9                  &12            & 1\\
  Design of architecture  &     15              &11                 & 16           &  3\\
  Development             &      3              & 4                 &  4           &  1\\
  No expectation          &     8               &5                  & 8            & 3\\
  Other                   &     6               &5                  & 7            & 1\\\hline
\end{tabular}
\vspace{-0.2in}
\end{table}

The adjusted Chi-square test confirms the independence and no association between the students' self-confidence levels and their expectations from the course, with $\chi^2$ of 3.4832, a p-value of  0.995, and Cramer V 0.00. We note, though, that the ANOVA test suggests that there is statistical evidence that the confidence level means of the perception categories are significantly different; the F-stat is 2.707, and the p-value is 0.07.

\subsection{Critical thinking}

\begin{table}[btp]
    \caption{Students' self-confidence levels vs the architecture aspects that they mentioned when differentiating the architectures of Web applications and IOT-based software.}
    \label{tab:WebIOTConfidence}

\centering
\rowcolors{2}{white}{lightgray}
\begin{tabular}{p{2.0in}p{0.9in}p{0.9in}p{0.9in}p{0.9in}p{1.0in} } 
\hline
\rowcolor{Gray}
Critical thinking              &Confident &Fair confidence &Moderate confidence &No confidence\\\hline
  Architecture drivers    &             17 &         9&              8  &           3\\
  Architecture knowledge  &                6 &       11&             13  &           1\\
  No definite answer      &                  3 &         8&         6&             2\\
  Patterns of the structure of the solutions  &      13 &              6&       20 &           3\\\hline
\end{tabular}
\vspace{-0.2in}
\end{table}

Table~\ref{tab:WebIOTConfidence} provides the frequencies of the students' critical thinking aspects vs. their self-confidence levels. We observe that the students who have high self-confidence discuss more the differences between Web-based and IOT-based applications in terms of architecture drivers and patterns of the solutions' structures and, to a lesser frequency, the architecture knowledge while the students who have moderate self-confidence discuss the differences in the patterns of the structure of the solutions and the architecture knowledge and, to lesser frequency, the differences in the architecture drivers between the two software types. Thus, we observe the students who have high, moderate, and fair self-confidence are mainly able to identify the differences in the architecture drivers, patterns of the solutions' structures, and architecture knowledge between the two architecture types and the students who did not express their critical thinking capability have mostly fair self-confidence. 

The Chi-square test confirms a dependency and weak association between the students' self-confidence levels and their critical thinking, with $\chi^2$ of 15.7898, a p-value of 0.063, and Cramer V 0.13. 

\subsection{Student' cognitive levels}

\begin{table}[btp]
  \caption{Relationship between cognitive levels and self-confidence levels.}
\label{tab:CognitiveConfidence}
\centering
\rowcolors{2}{white}{lightgray}
\begin{tabular}{p{2.0in}p{0.9in}p{0.9in}p{0.9in}p{0.9in}p{1.0in} } 
\hline
\rowcolor{Gray}
 Cognitive levels            &Confident &Fair confidence &Moderate confidence &No confidence\\\hline

      Applying  &           16    &          18     &             24       &     5\\
        Evaluating&            3  &             4   &                3  &           2\\
        Irrelevant &           5  &             4   &                5  &           1\\
        Understanding&        15  &             8   &               15  &           1\\\hline\hline
\end{tabular}
\end{table}

Table~\ref{tab:CognitiveConfidence} provides the frequencies of the students' cognitive levels vs. their self-confidence levels. We observe that the students who have applying cognitive levels have mostly moderate self-confidence levels, and the students who have understanding cognitive levels have high and moderate self-confidence levels. The paradox that high performers exhibit under-self-confidence is documented in other domains such as accounting ~\cite{RWW2012}. A possible reason is that the high performers know the limit of their abilities.  

The Chi-square test confirms the independence and no association between the students' self-confidence levels and their cognitive level, with $\chi^2$ of 5.7125, a p-value of 0.880, and Cramer V 0.00.  We note, though, that the ANOVA test suggests that there is statistical evidence that the confidence level means of the students' cognitive levels are significantly different; the F-stat is 5.05, and the p-value is 0.01.

%Note that the students who did not provide enough information to identify their cognitive levels have mostly high and moderate self-confidence levels. We looked at these students' responses and observed that they asked for more details to be able to answer the question related to the cognitive levels.

%The relationship between the cognitive levels and the students' self-confidence levels is expected: cognitive levels of the students do not discriminate the self-confidence levels. 

\subsection{Students' preferred learning methods.}

\begin{table}[btp]
     \caption{Students' self-confidence levels vs preferred learning methods.}
    \label{tab:learningmethodconfidence}

\centering
\rowcolors{2}{white}{lightgray}
\begin{tabular}{p{2.0in}p{0.9in}p{0.9in}p{0.9in}p{0.9in}p{1.0in} } 
\hline
\rowcolor{Gray}
Learning methods            &Confident &Fair confidence &Moderate confidence &No confidence\\\hline
 Active learning&         26   &           22   &               30 &            6\\
  Both approaches&          1  &             3  &                 5&             0\\
  Other           &         9  &             5  &                 7&             3\\
  Passive learning&         3  &             4  &                 5&             0\\ \hline
\hline\hline
\end{tabular}
\vspace{-0.2in}
\end{table}

%(5 have high self-confidence, 8 have moderate self-confidence, and 4 have fair self-confidence)

Table~\ref{tab:learningmethodconfidence} provides the frequencies of the students' preferred learning methods vs. their self-confidence levels. We observe that most of the students prefer active learn methods. The Chi-square test confirms the independence and no association between the students' self-confidence levels and their preferred learning methods, with $\chi^2$ of 6.1684, a p-value of 0.867, and Cramer V 0.00
We note, though, that the ANOVA test suggests that there is statistical evidence that the confidence level means of the students' cognitive levels are significantly different; the F-stat is 9.70, and the p-value is 0.001.

\section{Impacts and limitations of the study}\label{sec:limitations}

\begin{figure}[tbp]
    \centering
    \includegraphics[width=0.6\textwidth]{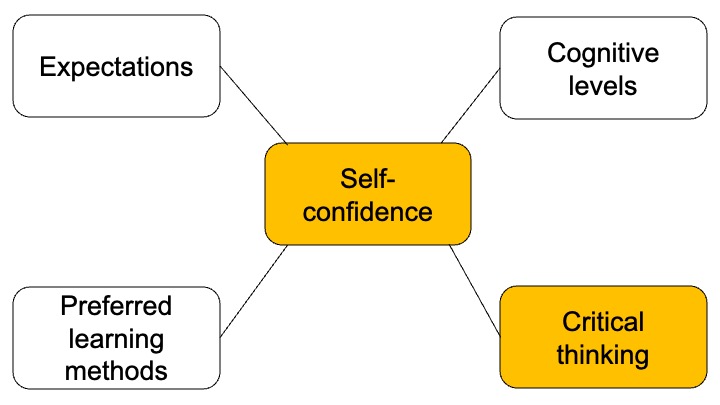}
\caption{Self-confidence and expected related factors. The yellow boxes indicates the variables associated with self-confidence.}\label{fig:summary} 
\label{fig:callGraphSample}\
\vspace{-0.2in}
\end{figure}

This paper explores a set of factors that we believe are related to undergraduate students' self-confidence levels, i.e., confidence in their abilities to design software architecture after taking a course on software architecture. The study found that the students' self-confidence is weakly associated with their critical thinking and does not depend on their cognitive levels, preferred learning methods, and perception about the course. Figure~\ref{fig:summary} depicts these relationships--the color indicates the associated factors. 

We reiterate that the students who have high cognitive levels did not have high self-confidence levels, and self-confidence is not associated with the cognitive levels.\footnote{We note that we cannot correlate the data with the students' assessment scores in the class because we did not request that in the  IRB before starting the study.}

We also did not see significant patterns from the analysis of the students' answers who do not have confidence in their ability to design software architecture. We found that these students have varying preferred learning methods (including practice), varying expectations from the class, and different cognitive levels. The results suggests that to improve the self-confidence of the students, the instructor should  work on improving the students' critical thinking capabilities. 

The main limitations of the study follow. First, we did not use a repeatable process to identify the factors that affect the students' self-confidence. The factors used in the study were identified in brainstorming sessions with colleagues: there would be other factors that impact the students' self-confidence that could be worth studying.

Second, the students provided their responses in text, and the authors coded the responses. One of the authors codes the students' response and the second authors validated the codes. Few codes were adjusted in the cases that were a disagreement between the coders is found. We acknowledge that the coders' perspective impacts the data extraction, which applies to qualitative research, in general. We, however, revisited the data extraction several times to reduce this limitation. We also cross-checked often the students' answers. 
The study shows that self-confidence is associated with the critical thinking of the students. This suggests that instructors can change their students' self-confidence by giving them knowledge about alternative solutions for solving given architecture problems, so they understand that there are conditions and implications of using architecture knowledge to solve architecture problems before asking them to apply architecture design methods~\cite{HOFMEISTER2007106}. 

\section{Conclusions}\label{sec:conclusions}

In this paper, the study analyzed the relationships between students' self-confidence levels and their expectations, preferred learning methods, cognitive levels, and  critical thinking. The study found that the students' self-confidence levels depend on their critical thinking capability but did not find dependency relationships between the self-confidence and students' cognitive levels or preferred learning methods. To improve the self-confidence of the students, the instructor should work on improving the students critical thinking capabilities. 

\section*{Acknowledgment}

The authors thank Yesdaulet Izenov for helping with the survey.

%\section*{References}
\bibliography{SoftArchEducation} 
\bibliographystyle{ieeetr}

%\begin{tcolorbox}

%\begin{enumerate}

%\item https://www.journals.elsevier.com/computers-and-education

%\end{enumerate}

%\end{tcolorbox}

\end{document}